\documentclass[12pt,preprint]{aastex}

\shorttitle{Protoplanetary Disk around HD~142527}
\shortauthors{Fujiwara et al.}

\begin{document}

\title{The Asymmetric Thermal Emission of Protoplanetary Disk Surrounding HD~142527 Seen by Subaru/COMICS\altaffilmark{1}}



\author{
Hideaki~Fujiwara\altaffilmark{2}, 
Mitsuhiko~Honda\altaffilmark{3}, 
Hirokazu~Kataza\altaffilmark{3}, 
Takuya~Yamashita\altaffilmark{4,2},
Takashi~Onaka\altaffilmark{2},
Misato~Fukagawa\altaffilmark{5,6},
Yoshiko~K.~Okamoto\altaffilmark{7}, 
Takashi~Miyata\altaffilmark{8}, 
Shigeyuki~Sako\altaffilmark{8}, 
Takuya~Fujiyoshi\altaffilmark{4},
and Itsuki~Sakon\altaffilmark{2}
}


\altaffiltext{1}{Based on data collected at Subaru Telescope, which is operated
by the National Astronomical Observatory of Japan.}
\altaffiltext{2}{Department of Astronomy, School of Science, University
of Tokyo, Bunkyo-ku, Tokyo 113-0033, Japan; fujiwara@astron.s.u-tokyo.ac.jp, onaka@astron.s.u-tokyo.ac.jp, isakon@astron.s.u-tokyo.ac.jp.}
\altaffiltext{3}{Institute of
Space and Astronautical Science, Japan Aerospace Exploration Agency, 3-1-1 Yoshinodai, Sagamihara, Kanagawa
229-8510, Japan; hondamt@ir.isas.jaxa.jp, kataza@ir.isas.ac.jp.}
\altaffiltext{4}{Subaru Telescope, National Astronomical Observatory of
Japan, 650 North A'ohoku Place, Hilo, HI 96720;
takuya@subaru.naoj.org, tak@subaru.naoj.org.}
\altaffiltext{5}{{\it Spitzer} Science Center, California Institute of Technology, Mail Code 220-6, Pasadena, CA 91125, U.S.A.; 
misato@ipac.caltech.edu.}
\altaffiltext{6}{
Nagoya University, Furo-cho, Chikusa-ku, Nagoya 464-8602, Japan.}
\altaffiltext{7}{Institute of Astrophysics and Planetary Sciences,
Ibaraki University, Bunkyo 2-1-1, Mito, Ibaraki 310-8512, Japan; yokamoto@mx.ibaraki.ac.jp.}
\altaffiltext{8}{Institute of Astronomy, University of Tokyo, 2-21-1 Osawa, Mitaka, Tokyo 181-0015, Japan; 
tmiyata@ioa.s.u-tokyo.ac.jp, sako@ioa.s.u-tokyo.ac.jp.}


\begin{abstract}
Mid-infrared (MIR) images of the Herbig Ae star HD~142527 were obtained at 18.8 and 24.5~$\micron$ with the Subaru/COMICS. Bright extended arc-like emission (outer disk) is recognized at $r=0\farcs85$ together with a strong central source (inner disk) and a gap around $r=0\farcs6$ in the both images. Thermal emission of the eastern side is much brighter than that of the western side in the MIR.  We estimate the dust size as a few~$\micron$ from the observed color of the extended emission and the distance from the star. The dust temperature $T$ and the optical depth $\tau$ of the MIR emitting dust are also derived from the two images as $T=82\pm 1$~K, $\tau=0.052\pm 0.001$ for the eastern side and $T=85\pm 3$~K, $\tau=0.018\pm0.001$ for the western side. The observed asymmetry in the brightness can be attributed to the difference in the optical depth of the MIR emitting dust. To account for the present observations, we propose an inclined disk model, in which the outer disk is inclined along the east-west direction with the eastern side being in the far side and the inner rim of the outer disk in the eatern side is exposed directly to us. The proposed model can successfully account for the MIR observations as well as near-infrared (NIR) images of the scattering light, in which the asymmetry is seen in the opposite sense and the forward scattering light (near side -- western side) is brighter.
\end{abstract}


\section{Introduction}

High angular resolution, multiwavelength imaging of protoplanetary
disks can provide a unique insight into planet formation processes,
since detailed structure of a disk should bear traces of
disk formation history and of young or mature planets.
Mid-infrared (MIR) observations especially trace the thermal radiation from warm dust, 
which allow us to examine the temperature structure and the dust distribution quantitatively.
They also allow us to discuss the structure near the central star because the contribution of the photosphere is relatively small and coronagraphic observations are generally not required for the MIR.

HD~142527 (F7IIIe) is classified as a Herbig Ae star as confirmed by its spectrum and photometry, which exhibit the typical characteristics of a pre-main sequence (PMS) star \citep{Waelkens96}. {\it Hipparcos} measurements suggest its distance of $d=200^{+60}_{-40}$~pc 
\citep{vandenAncker98}.
The stellar age and mass are $1$~Myr and $2.5 \pm 0.3 M_{\odot}$, respectively, derived by comparing the position of HD~142527 in the HR diagram to theoretical PMS tracks \citep{vanboekel05}.
\cite{malfait98} reported that infrared excess of HD~142527 is comparable to the stellar luminosity, suggesting that a large amount of cold material is present around the star. Radio observations show the disk mass as $M_{\rm disk}=0.15M_\odot$ \citep{acke04}, which is more massive than a typical disk mass by an order of the magnitude. 
\cite{vanboekel04} reported the central concentration of crystalline silicate in the disk.

\cite{fukagawa06} resolved an almost face-on disk around HD~142527 in the near-infrared (NIR) scattered light using the Subaru/CIAO. They discovered arc-like components facing each other along the east-west direction (``banana-split structure'') with one spiral arm extending to the north from the western arc in the disk.
From this unusual morphology, the presence of an unseen eccentric binary and a recent stellar encounter are suggested.
However, the coronagraphic mask made it difficult to quantitatively discuss the spatial structure in the vicinity of the central star solely from scattered light observations.

In this letter, we present high spatial resolution MIR images of HD~142527 obtained with the Subaru Telescope.
The obtained images provide significant information on the nature of the protoplanetary disk around HD~142527.

\section{Observations and data analysis}

We observed the Herbig Ae star HD~142527 with the COoled Mid-Infrared Camera
and Spectrometer \citep[COMICS;][]{kataza00,okamoto02,sako03} mounted on
the 8.2~m Subaru Telescope on 2004 July 13 and 2005 August 24. 
Imaging observations with the 
18.75~$\micron$ ($\Delta\lambda$=0.9~$\micron$) and 24.56~$\micron$ ($\Delta\lambda$=0.75~$\micron$) bands were carried out.
The pixel scale was $0\farcs13$~pixel$^{-1}$.
To cancel out the background radiation,
the secondary mirror chopping was used at a frequency of 0.45 Hz for 18.8~$\micron$ and 0.46 Hz for 24.5~$\micron$ with a $10\arcsec$ throw. For the Subaru/COMICS, the residual pattern in the chopping
subtraction is very small and negligible compared to the brightness
of the target in most cases. We did not apply nodding for the present
observations. We selected standard stars (HD~136422 for 18.8~$\micron$ and HD~146051 for 24.5~$\micron$) from \cite{cohen99} as flux calibrators and reference point-spread functions (PSFs).
We observed them before or after the observations of the target in the same manner.
The observation parameters are summarized in Table~\ref{obslog}.

For the data reduction, we used our own reduction tools and IRAF.
The standard chopping pair subtraction and the shift-and-add method were employed.
We also corrected the effect of the different airmass between the object and
the standard star by estimating the difference in atmospheric absorption using 
the ATRAN software \citep{lord92}.
The derived total absolute fluxes are $14.6\pm0.2$~Jy and $19.3\pm0.1$~Jy at 18.8 and 24.5~$\micron$, respectively, and they were consistent with the flux of the ISO-SWS spectrum \citep{sloan03}.

\section{Results}
\label{result}

\begin{figure}
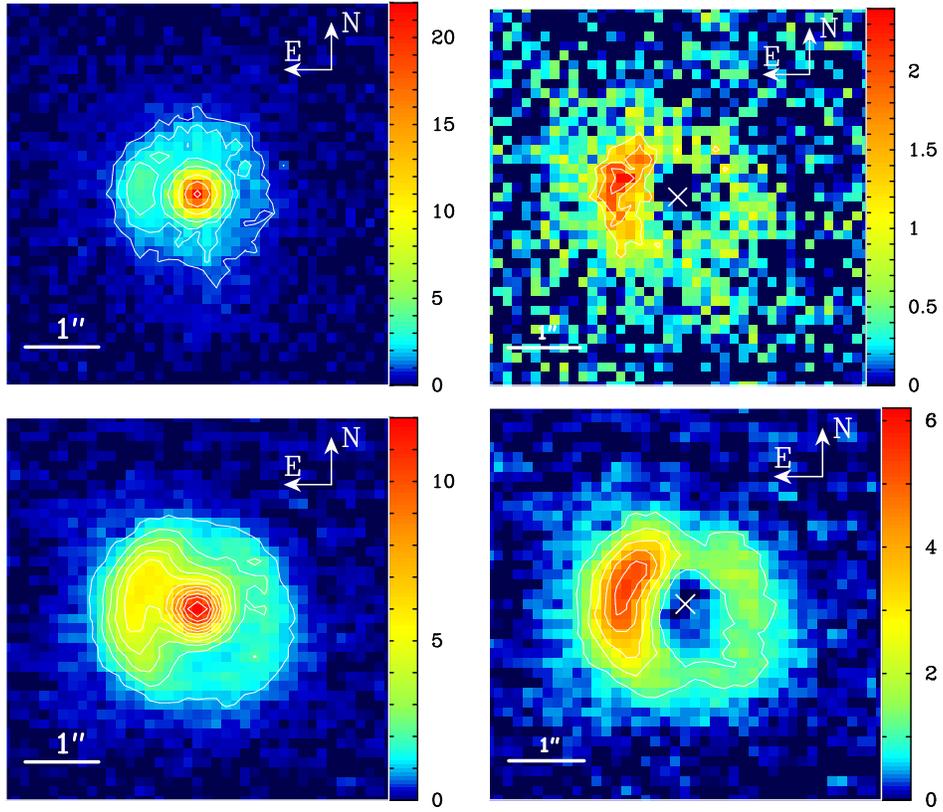

  \begin{center}
    \begin{tabular}{cc}
      \resizebox{60mm}{!}{\includegraphics{HD142527Q188.eps}} &
      \resizebox{60mm}{!}{\includegraphics{HD142527Q188.subPSF.eps}} \\
      \resizebox{60mm}{!}{\includegraphics{HD142527Q245.eps}} &
      \resizebox{60mm}{!}{\includegraphics{HD142527Q245.subPSF.eps}} \\
    \end{tabular}
    \caption{Observed images of HD~142527 (left) and central-component-subtracted images of HD~142527 (right). 
	These images are in 18.8~$\micron$ (top) and 24.5~$\micron$ (bottom). 
	The color bars in the right are in the units of Jy arcsec$^{-2}$.\label{Qimg}}
  \end{center}
\end{figure}

\begin{figure}
\epsscale{0.5}
\plotone{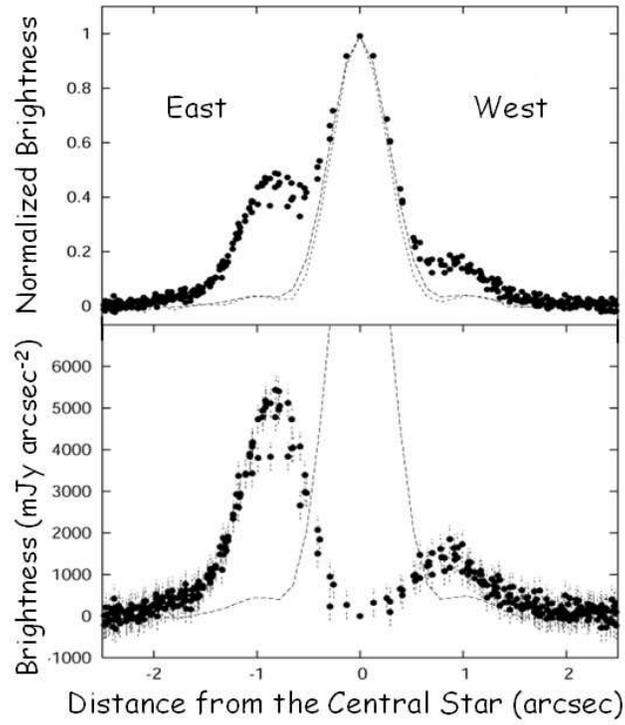}
\caption{24.5~$\micron$ radial profiles of the eastern (PA=$60-120^\circ$) and western (PA$=240-300^\circ$) side. {\it Top}: pre-subtraction of the central component; {\it Bottom}: post-subtraction. The profiles of the subtracted component and PSF are also plotted with the dashed and dotted lines, respectively.\label{Q245_EW_pro}}
\end{figure}

The obtained images are shown in Fig.~\ref{Qimg}~({\it left}).
A prominent extended radiation component is recognized in addition to a central compact component both in the 18.8 and 24.5~$\micron$ images. By comparing with the standard stars, the fluxes of the central component of HD~142527 are measured as $10.4\pm0.08$~Jy and $6.7\pm0.05$~Jy at 18.8 and 24.5~$\micron$, respectively.
The central component is thought to be the thermal radiation from hot dust within the inner disk close to the star because the contribution from the photosphere is less than 0.1~Jy and it is much smaller than the observed emission at these wavelengths. 
Fig.~\ref{Q245_EW_pro}~({\it top}) shows the profile of 24.5~$\micron$ image along the east-west direction.
While the profile of the central component ($r \le 0\farcs5$) at 18.8~$\micron$ is consistent with the PSF (FWHM$=0\farcs52$), that at 24.5~$\micron$ is broader than the PSF (FWHM$=0\farcs61$). 
The convolved PSF by the Gaussian of FWHM$= 0\farcs28$ gave the best fit to the profile of the central component at 24.5~$\micron$.
This suggests that the inner disk may be extended and its size can be estimated to be $\ge 80$~AU from the brightness of the central component at 24.5~$\micron$ simply assuming that the disk is a uniform blackbody.
Since N-band spectroscopic observations of HD~142527 by \cite{vanboekel04} show broad silicate features, $\micron$-sized large grains are dominant in the inner disk.
From the fluxes at 18.8 and 24.5~$\micron$, the temperature and optical depth of the MIR emitting dust within the inner disk are derived as $T=293$~K and $\tau=4.4 \times 10^{-4}$, respectively, by assuming a single size distribution of silicate for simplicity and by using the emissivity of 2.5~$\micron$-sized astronomical silicate dust \citep{laor93}.

The extended component is thought to be the thermal radiation from the material of relatively low temperatures ($T<150$~K) in the outer region of the protoplanetary disk since it is brighter at 24.5~$\micron$ than at 18.8~$\micron$.
In order to examine the extended emission in detail, we subtracted the central component.
For the 18.8~$\micron$ data, we subtracted the scaled reference PSF as the central component from the image of HD~142527.
For the 24.5~$\micron$ data, we subtracted the reference PSF that is convolved by a Gaussian of FWHM$=0\farcs28$ as the central component since the central component may be slightly extended (see above).
The reference PSFs are very stable and the uncertainties of the PSF subtractions are estimated as 2.6~mJy arcsec$^{-2}$ at 24.5~$\micron$ around the diffraction ring even in the worst case and less than the background noise level of 6.7~mJy arcsec$^{-2}$ at 18.8~$\micron$. The uncertainties are included to the errors of the following results.

\begin{figure}
\plotone{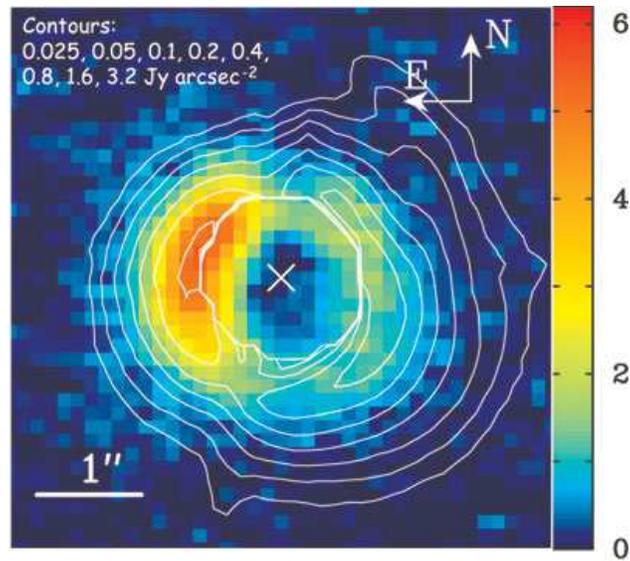}
\caption{Central-component-subtracted 24.5~$\micron$ image of HD~142527 (color) and $H$-band scattered light image with CIAO (contour). The color bar in the right is in the units of Jy arcsec$^{-2}$. \label{Q_CIAO}}
\end{figure}

After the subtraction of the central component, asymmetric ring-like structures are clearly seen (Fig.~\ref{Qimg}~{\it right}). The eastern side is much brighter than the western side both at 18.8 and 24.5~$\micron$. We compare the MIR and NIR images in Fig.~\ref{Q_CIAO}. They were aligned with each other by adjusting the stellar position. The stellar position in the MIR image was detemined as the peak position of the central component derived by Gaussian fitting and its uncertainty is about 0.1~pixel$=0\farcs013$. The maximum offset between the position of the MIR and NIR emission is $0\farcs013$. 
The observed extended emission arcs are also seen in the NIR image as ``banana-split structure'' \citep{fukagawa06}. 
Gaps in the north and south are also clearly seen both in the MIR and NIR images.
However, the contrast of the two arcs is quite different from the NIR scattered light images, in which the eastern side is brighter than the western side. 
The spiral arm observed in the NIR is not detected in our MIR images.

To discuss the azimuthal variation in the extended emission, 
we divided the observed extended emission into two regions as the position angle (PA) $=0-180^\circ$ and $180-360^\circ$; eastern and western regions. 
The radial profiles of the eastern and western sides are shown in Fig.~\ref{Q245_EW_pro}.
The eastern side is about 3 times brighter than the western side.
The radial profiles are almost Gaussian-like and the peaks are at about $r=0\farcs85$ ($=170$~AU) for both sides. The peak position and profile width (FWHM) in each region derived by Gaussian fitting for the 24.5~$\micron$ are listed in Table~\ref{derived_parameters}. The widths of both arcs are broader than the beam size.

\begin{figure}
\epsscale{0.5}
\plotone{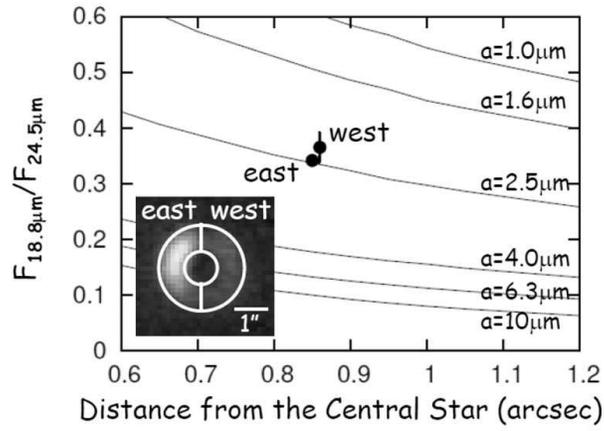}
\caption{Color and position of the eastern and western arc. The color profiles calculated by using the emissivity of astronomical silicate are also plotted with the solid lines. \label{rc_region_east-west}}
\end{figure}

By integrating the signals within the range of $0\farcs5 \le r \le 1\farcs2$, we derived the fluxes of the eastern and western regions at 18.8 and 24.5~$\micron$ and the flux ratio of $F_{18.8\micron}/F_{24.5\micron}$ as a color indicator of the dust emission (Table~\ref{derived_parameters}).
We calculated the predicted color profiles of the MIR emitting dust around HD~142527 for the emissivities of various sized astronomical silicate \citep{laor93} and compared them with the observed colors (Fig.~\ref{rc_region_east-west}).
We assumed $L_*=69L_\odot$ and $T_{\rm eff}=6250$~K for the photospheric spectrum of HD~142527 \citep{vandenAncker98}.
Fig.~\ref{rc_region_east-west} shows that the observed colors of both regions are in agreement with the model calculation for 2.5~$\micron$-sized silicate. Thus we can suggest the typical size of the MIR emitting dust within the outer disk of HD~142527 as a few~$\micron$ when we assume dust species as amorphous silicate. 
\cite{fukagawa06} also argue that the size of grains responsible for the scattering is $\gtrsim 1~\micron$ because the color of scattered light from dust around HD~142527 is similar to the color of the star, being in agreement with ours.
The dust grains around HD~142527 are larger than the grains in the interstellar medium, suggesting that the dust grains in the disk have already grown to some extent due to coagulation of smaller dust particles.
We also derived the temperature and optical depth of the MIR emitting dust within each region from the fluxes of 18.8~$\micron$ and 24.5~$\micron$, assuming a single size distribution of 2.5~$\micron$-sized silicate \citep{laor93}.
We derived the temperatures as $T=82 \pm 1$~K and $T=85\pm3$~K and the optical depths of the MIR emitting dust as $\tau=0.057\pm0.001$ and $\tau=0.018 \pm 0.001$ for the eastern and western regions, respectively.
While the temperatures of the eastern and western regions are in agreement with each other within the errors, the optical depth of the eastern region is 3 times larger than that of the western region.
The asymmetry in the brightness of the extended emission originates from the difference of the optical depth between the eastern and western side.

While the above results are derived on the assumption of $L_*=69L_\odot$, \cite{vanboekel05} proposed $L_*\sim30L_\odot$.
Taking this luminosity, then we obtain 1.6~$\micron$ as a typical size of grains.
The effect of the radiation from the inner disk has also been examined by assuming that 
the inner disk emits as a blackbody of $T=1200$~K and $L\sim0.5L_*$, which is estimated from the spectral energy distribution (SED).
The estimated dust size becomes 2~$\micron$.
Thus we conclude that the presence of $\micron$-sized grains in the disk is secure.

\section{Discussion}

As a global distribution of the material in the disk around HD~142527, we suggest that there are a large amount of cold materials in the outer region of $r \gtrsim 170$AU (outer disk), while there are quite a few materials present in the inner region. 
Since we can see a drop of radiation at $r=0\farcs6$ in the images, a gap in the material density must exist.
The existence of the material gap is compatible with the MIR flux gap seen in the SED \citep{acke04}.
\cite{leinert04} suggested the presence of ``a nearby very red source'' whose temperature is $T \sim 70$~K as the origin of the large far-infrared excess seen in the SED.
However, our observations demonstrate that the large far-infrared excess is attributed to the outer disk associated with the star, not to ``a nearby very red source''.

\begin{figure}
\epsscale{0.5}
\plotone{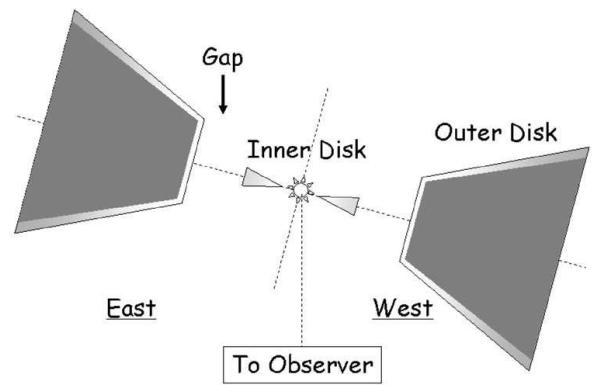}
\caption{A schematic view of the model of the HD~142527 system. The gap between the inner and outer disk exists at $r=80-170$~AU. \label{rim_model}}
\end{figure}

As we mentioned in \S \ref{result}, the asymmetric brightness of the outer disk of HD~142527 can be attributed to the difference in $\tau$ of the MIR emitting dust. 
NIR scattered light observations of HD~142527 by CIAO show that the western side of the extended disk is brighter than the eastern side by 2.0 mag \citep{fukagawa06}.
850~$\micron$ continuum observations by Submillimeter Array (SMA), which is thought to be optically thin to dust emission and a good tracer of the total amount of dust, show an east-west symmetric structure of the extended disk (Ohashi et al, in prep.).
To explain these observations, we propose a model that the dust disk is geometrically thick and inclined as the eastern side is farther to us and the inner rim of the eastern side of the outer disk is exposed to us.
Fig.~\ref{rim_model} shows a schematic view of the model of the HD~142527 system.
Since the inner edge and rim are hottest within the outer disk and affect the MIR brightness most strongly, the emitting region of the outer disk appears like a narrow ring. 
Therefore the optical depth we mentioned above does not directly correspond to the vertical depth of the MIR emitting layer of the disk, but rather indicates the area size of the MIR emitting region within the beam.
If the disk is inclined as the eastern side is farther to us, the eastern inner rim of the outer disk can be seen directly while the western one cannot because the cooler materials farther out are obscuring the inner hot dust. 
We suggest that the difference of visibility of the inner rim makes the strong asymmetric MIR brightness of the outer disk.
The model we are proposing here can also explain that the western side is brighter in the NIR scattered light \citep{fukagawa06}.
In the scattered light, we observe back scattered light in the eastern side and forward scattered light in the western side.
For grains those sizes are similar to or larger than the wavelength in question, the forward scattering is much stronger than the back scattering, resulting in the observed asymmetry in the NIR.
An axial symmetric disk inclined to the east-west direction is also consistent with the symmetric structure of the extended disk traced by 850~$\micron$ radiation.
It is not necessary to consider an axial asymmetric material distribution such as a circumbinary disk to account for the observations.

Details of the appearance of a protoplanetary disk, including a ``peculiar horseshoe-like feature'' for models with an inclination of 30$^\circ$, were predicted in simulations of \cite{kessel98}. A model 12~$\micron$ intensity map in their paper shows a morphology strikingly similar to our images of the outer disk of HD~142527.
NIR interferometric observations of LkH$\alpha$~101 show that an asymmetric arc-like radiation component is present in addition to a central compact component \citep{tuthill02}.
In the case of LkH$\alpha$~101, the asymmetric arc-like structure is thought to be attributed to the 
hot dust within the inner rim of the inner hole made by dust sublimation.
Though the observed spatial scale and wavelength are different, HD~142527 and LkH$\alpha$~101 agree on the asymmetric arc-like morphology of thermal emission.
In the case of HD~142527, photoevaporation of dust can be ruled out as the cause of the gap because the position of gap is far from the central star. A secondary object, if any, may be able to create the gap. 
High spatial resolution radio observations will be required in order to discuss the origin of the gap in detail.

\acknowledgments

We appreciate the support from the Subaru Telescope staff. 
We also thank Dr.~N.~Ohashi for providing the SMA data and Dr.~Y.~Nakada and Dr.~C.~Waelkens for useful comments.
This research was supported by the MEXT, ``Development of Extra-solar Planetary Science''.
M.~H., M.~F. and I.~S. are supported by the JSPS.

\clearpage

\clearpage

\clearpage

\clearpage

\clearpage

\clearpage

\begin{deluxetable}{lllrrl}
\tabletypesize{\scriptsize}
\tablecaption{
Summary of the observations of HD~142527  \label{obslog}
}
\tablewidth{0pt}
\tablehead{
\colhead{Object} & \colhead{Filter}   & \colhead{Date[UT]}   &
\colhead{Integ.[sec]} & \colhead{AirMass} & \colhead{Comments}
}
\startdata
HD~142527    & Q18.8 & 2004/7/13 &  502    & 2.1  & \nodata \\
HD~136422  & Q18.8 & 2004/7/13 &  201    & 1.8  & standard \\
HD~142527    & Q24.5 & 2005/8/24 &  1404   & 2.3-2.4  & \nodata \\
HD~146051  & Q24.5 & 2005/8/24 &  201    & 1.2  & standard \\
\enddata
\end{deluxetable}

\clearpage

\begin{deluxetable}{lrrcclrl}
\tabletypesize{\scriptsize}
\tablecaption{
Derived physical parameters for each region
\label{derived_parameters}}
\tablewidth{0pt}
\tablehead{
\colhead{Region} & \colhead{F18.8[Jy]} & \colhead{F24.5[Jy]} & \colhead{Position} & \colhead{Width} & \colhead{F18.8/F24.5} & \colhead{T[K]} & \colhead{$\tau_{24.5~\micron}$} 
}
\startdata
Center & 10.4 $\pm$ 0.08 & 6.7 $\pm$ 0.05 & \nodata & \nodata & $1.55 \pm 0.02$ & $293 \pm 8$ & $ 4.4 \times 10^{-4}
$ \\
East & 1.98 $\pm$ 0.05 & 5.52 $\pm$ 0.06 & 0\farcs85 & 0\farcs87 & $0.34 \pm 0.01$ & $82 \pm 1$ & $0.057 \pm 0.001$ \\
West & 0.77 $\pm$ 0.05 & 2.12 $\pm$ 0.06 & 0\farcs86 & 0\farcs94 & $0.37 \pm 0.03$ & $85 \pm 3$ & $0.018 \pm 0.001$ \\
\enddata
\end{deluxetable}

\end{document}